\documentstyle[12pt]{article}

\catcode`\@=11
\long\def\@makefntext#1{
\protect\noindent \hbox to 3.2pt {\hskip-.9pt
$^{{\ninerm\@thefnmark}}$\hfil}#1\hfill}		

\def\@makefnmark{\hbox to 0pt{$^{\@thefnmark}$\hss}}  
\def\ps@myheadings{\let\@mkboth\@gobbletwo
\def\@oddhead{\hbox{}
\rightmark\hfil\ninerm\thepage}
\def\@oddfoot{}\def\@evenhead{\ninerm\thepage\hfil
\leftmark\hbox{}}\def\@evenfoot{}
\def\sectionmark##1{}\def\subsectionmark##1{}}

\renewcommand{\thefootnote}{\fnsymbol{footnote}}

\newcounter{sectionc}\newcounter{subsectionc}\newcounter{subsubsectionc}
\renewcommand{\section}[1] {\vspace*{0.6cm}\addtocounter{sectionc}{1}
\setcounter{subsectionc}{0}\setcounter{subsubsectionc}{0}\noindent
	{\normalsize\bf\thesectionc. #1}\par\vspace*{0.4cm}}
\renewcommand{\subsection}[1] {\vspace*{0.6cm}\addtocounter{subsectionc}{1}
	\setcounter{subsubsectionc}{0}\noindent
	{\normalsize\it\thesectionc.\thesubsectionc. #1}\par\vspace*{0.4cm}}
\renewcommand{\subsubsection}[1]
{\vspace*{0.6cm}\addtocounter{subsubsectionc}{1}
	\noindent {\normalsize\rm\thesectionc.\thesubsectionc.\thesubsubsectionc.
	#1}\par\vspace*{0.4cm}}

\newcounter{appendixc}
\newcounter{subappendixc}[appendixc]
\newcounter{subsubappendixc}[subappendixc]

\renewcommand{\appendix}[1] {\vspace*{0.6cm}
        \refstepcounter{appendixc}
        \setcounter{figure}{0}
        \setcounter{table}{0}
        \setcounter{equation}{0}
        \renewcommand{\thefigure}{\Alph{appendixc}.\arabic{figure}}
        \renewcommand{\thetable}{\Alph{appendixc}.\arabic{table}}
        \renewcommand{\theappendixc}{\Alph{appendixc}}
        \renewcommand{\theequation}{\Alph{appendixc}.\arabic{equation}}
        \noindent{\bf Appendix \theappendixc #1}\par\vspace*{0.4cm}}

\def\abstracts#1{{
\centering{\begin{minipage}{12.2truecm}\vspace*{.1cm}
        \footnotesize\baselineskip=12pt\noindent
	\parindent=0pt #1
	\end{minipage}}\par}}

\renewenvironment{thebibliography}[1]
	{\begin{list}{\arabic{enumi}.}
	{\usecounter{enumi}\setlength{\parsep}{0pt}
\setlength{\leftmargin 1.25cm}{\rightmargin 0pt}
	 \setlength{\itemsep}{0pt} \settowidth
	{\labelwidth}{#1.}\sloppy}}{\end{list}}

\newcounter{itemlistc}
\newcounter{romanlistc}
\newcounter{alphlistc}
\newcounter{arabiclistc}

\newcommand{\fcaption}[1]{
        \refstepcounter{figure}
        \setbox\@tempboxa = \hbox{\footnotesize Fig.~\thefigure. #1}
        \ifdim \wd\@tempboxa > 6in
           {\begin{center}
        \parbox{6in}{\footnotesize\baselineskip=12pt Fig.~\thefigure. #1}
            \end{center}}
        \else
             {\begin{center}
             {\footnotesize Fig.~\thefigure. #1}
              \end{center}}
        \fi}

\newcommand{\tcaption}[1]{
        \refstepcounter{table}
        \setbox\@tempboxa = \hbox{\footnotesize Table~\thetable. #1}
        \ifdim \wd\@tempboxa > 6in
           {\begin{center}
        \parbox{6in}{\footnotesize\baselineskip=12pt Table~\thetable. #1}
            \end{center}}
        \else
             {\begin{center}
             {\footnotesize Table~\thetable. #1}
              \end{center}}
        \fi}

\def\@citex[#1]#2{\if@filesw\immediate\write\@auxout
	{\string\citation{#2}}\fi
\def\@citea{}\@cite{\@for\@citeb:=#2\do
	{\@citea\def\@citea{,}\@ifundefined
	{b@\@citeb}{{\bf ?}\@warning
	{Citation `\@citeb' on page \thepage \space undefined}}
	{\csname b@\@citeb\endcsname}}}{#1}}

\newif\if@cghi
\def\cite{\@cghitrue\@ifnextchar [{\@tempswatrue
	\@citex}{\@tempswafalse\@citex[]}}
\def\citelow{\@cghifalse\@ifnextchar [{\@tempswatrue
	\@citex}{\@tempswafalse\@citex[]}}
\def\@cite#1#2{{$\null^{#1}$\if@tempswa\typeout
	{IJCGA warning: optional citation argument
	ignored: `#2'} \fi}}

 1
 1
 1

\font\ninerm=cmr9

\begin{document}

\begin{flushright}
UUITP-11/95
\end{flushright}

\centerline{\normalsize\bf EFFECTIVE MODELS}
\baselineskip=22pt
\centerline{\normalsize\bf OF THE ELECTROWEAK PHASE TRANSITION\footnote{
To appear in the Procs. of PASCOS/HOPKINS 1995 Symposium, Baltimore, March
22-25, 1995 }}

\centerline{\footnotesize ANDR\'AS PATK\'OS}
\baselineskip=13pt
\centerline{\footnotesize\it Department of Atomic Physics, E\"otv\"os
University}

\centerline{\footnotesize\it Puskin u. 5-7., H-1088 Budapest Hungary}

\vspace*{0.9cm}
\abstracts{The consecutive integration over the distinct  mass scales ${\cal
O}(T),{\cal O}(gT)$ leads to a hierarchy of effective models for the
electroweak phase transition. Different techniques for the realisation of such
strategy are reviewed. Advantages and difficulties resulting from the use of
reduced models are discussed.}

\normalsize\baselineskip=15pt
\setcounter{footnote}{0}
\renewcommand{\thefootnote}{\alph{footnote}}

\section{Introduction}
Infrared  improved perturbative treatments of the electroweak phase transition
\cite{huet,buch1,zwirner,arnold1,fodor1}have reinforced the pioneering
suggestion of Kirzhnits and Linde\cite{linde1}, that "massless" finite
temperature magnetic fluctuations might drive this transition into the first
order regime. The effect of the out-of-equilibrium state associated with a
discontinous transition might be essential for understanding the possible
origin and the survival of a cosmological B--L asymmetry \cite{shapo1,arnold}.

The question of existence of a characteristic mass scale of finite temperature
non-Abelian magnetic fluctuations has been discussed extensively in the past
decade, by means of numerical \cite{MC1,MC2,MC3} and semi-classical \cite{biro}
techniques, and with help of self-consistent coupled Schwinger--Dyson equations
\cite{buch1,zwirner,buch2,philip}.
Also, it has been argued that several finite temperature quantities (e.g.
electric screening mass, etc.) would be sensitive to the existence of a
non-zero magnetic scale \cite{rebhan}.
All approximate studies conjecture the magnetic scale to be of
${\cal O}(g^2T)$, ($g$ being the gauge coupling).
More generally, it is expected that phenomena which are perturbatively not
accessible  in finite temperature Higgs models would occur at this scale.

Therefore with no hesitation one can integrate out of the partition function of
the finite temperature electroweak theory all non-static Matsubara modes, since
they are characterised by the high momentum scale $2\pi T$. This is the
background for the application of the "conventional" reduction program, which
has been realised for the Higgs-systems to date at 1-loop level
\cite{farak1,jako,jako2}. In the course of the integration the static parts of
the electric components of the vector fields, $A_0$ and of the Higgs-doublet
receive thermal mass contributions ${\cal O}(gT)$.

In this way the $A_0$-field (at least in the weak coupling limit) defines a new
scale which is still distinctly bigger than the magnetic one, therefore it can
also be integrated out of the theory. It is important to emphasize that the
elimination of the static $A_0$ is justifiable only in the present hierarchical
approach. The experimental value of $g\sim2/3$, however might raise doubts
about the quantitative accuracy of another 1-loop approximation
\cite{jako,farak2}, even though in terms of the original theory it corresponds
to an infinite resummation. More accurate integration schemes (for instance,
Renormalisation Group improved 1-loop integration \cite{patkos}) might be
invoked in order to estimate the systematic errors introduced by simpler
approximations.

A similar integration over the Higgs fields is not advisable, especially in the
temperature range of the phase transition, where the thermal contribution tends
to compensate the wrong sign squared mass, defining the theory at $T=0$.

In section 2, I summarize 1-loop results for reduced theories of the
electroweak phase transition.
In section 3 the strategy of non-perturbative (lattice MC) investigations of
the effective theories will be outlined. Its application will be illustrated on
the example of a pure scalar (order--parameter) effective theory of the
electroweak phase transition \cite{karsch}.
The effects of non-renormalisable and of non-local operators, appearing when
more accurate mapping of the full theory on a 3 dimensional effective model
is attempted, will be discussed in section 4 on the example of the finite
temperature O(N) symmetric scalar model \cite{jako3}. In particular, I shall
compare the strategy of consecutive integrations to the "matching" program of
Braaten and Nieto \cite{braaten}.

\section{Hierarchy of effective theories at 1-loop level}

The object of study in most investigations is the Euclidean version of the
SU(2) Higgs model:
\begin{equation}
L[A_\mu ,\phi ]={1\over 4}F_{mn}^aF_{mn}^a+{1\over 2}(D_m\phi )^\dagger
(D_m\phi )+{1\over 2}m^2\phi^\dagger\phi+{1\over 24}(\phi^\dagger\phi )^2
+L_{c.t.}^{4D},
\end{equation}
with $F_{mn}^a=\partial_mA_n^a-\partial_{n}A_m^a+g\epsilon^{abc}A_m^bA_n^c,
{}~D_m\phi =(\partial_m+ig\tau^aA_m^a/2)\phi$. $A_m^a$ is the 4-dimensional
vector field transforming as a triplet under the SU(2) gauge group, while
$\phi$ is a complex Higgs-doublet. $L_{c.t.}^{4D}$ refers to the temperature
independent counterterms absorbing ultraviolet singularities of the
4-dimensional theory. One expects that once controlling the phase transition of
this simplified system one will be able to handle the complete SU(2) x U(1)
theory. The inclusion of chiral fermions, however, is yet posing insurmountable
difficulties to lattice studies of the full 4-dimensional finite temperature
model. In this respect, the integration over all fermionic modes (characterised
by the mass scale $\pi T$) is the only practical solution.

\subsection{Integration of non--static modes}

Complete 1-loop integration has been performed in the background of static
$A_i(x), A_0(x),\phi (x)$ fields in general covariant gauge \cite{jako2} and
also in Landau gauge \cite{farak2}. With appropriate (gauge parameter
dependent) field-, mass- and coupling constant renormalisations one arrives at
the following gauge independent effective (reduced) system in 3 dimensions:
\begin{eqnarray}
&
L_{eff}^{I}={1\over 4}F_{ij}^aF_{ij}^a+{1\over
2}(D_i^{adj}A_0)^a(D_i^{adj}A_0)^a+{1\over 2}(D_i\phi )^\dagger(D_i\phi
)\nonumber\\
&
+{1\over 2}m_3^2\phi^\dagger\phi+{1\over 2}m_D^2(T)(A_0^a)^2\nonumber\\
&
+{1\over 2}g_3^2(A_0^a)^2\phi^\dagger\phi+{1\over 24}\lambda_3(\phi^\dagger\phi
)^2+{17g_3^4\over 192\pi^2T}[(A_0^a)^2]^2+{\rm higher~ dim.~ op's},
\end{eqnarray}
where
\begin{eqnarray}
&
g_3^2=g_R^2T, ~\lambda_3=\lambda_R T,~m_D^2(T)={5\over 6}g_3^2T,\nonumber\\
&
m_3^2=m_R^2+({3\over 16}g_3^2+{1\over 12}\lambda )T,
\label{3dcoup}
\end{eqnarray}
($(D_i^{adj}A_0)^a=(\partial_i\delta^{ac}+g_3\epsilon^{abc}A_i^b)A_0^c$). The
quantities with index "R" above refer to couplings renormalised in 4
dimensional sense (this index will be omitted below). The omission of higher
dimensional operators from the effective action will be discussed in section 4.

Counterterms of this 3 dimensional system cannot be chosen freely: the
procedure of the "projection" of the full system onto the static variables
should induce those singularities of the action which are necessary for
ensuring the finiteness of the static effective theory. Even those divergences
which would arise if higher dimensional (usually called non--renormalisable)
operators would be retained in the effective action should be canceled by
appropriate induced counterterms. In principle, therefore, it presents no
particular interest to distinguish between approximate versions of the
effective theories which are renormalisable on their own in 3-dimensional sense
and those which are not. At 1-loop level of the reduction, however, only mass
counterterms are produced ( with a sharp momentum cut-off):
\begin{equation}
L_{c.t.}^{I}=-{1\over 2}\phi^\dagger\phi({9\over
4}g_3^2+\lambda_3){\Lambda\over 2\pi^2}-{1\over 2}(A_0^a)^25g_3^2{\Lambda\over
2\pi^2}.
\end{equation}

\subsection{Integration of the SU(2) triplet field $A_0^a$}

Further integration over fields characterised by the scale ${\cal O}(gT)$ has
been proposed in \cite{farak2,jako}. 1-loop integration can easily be performed
(there is no need for fixing any gauge) resulting in a variant of the 3D Higgs
model:
\begin{eqnarray}
&
L_{eff}^{II}={1\over 4\mu}F_{ij}^aF_{ij}^a+{1\over 2}(D_i\phi)^\dagger(D_i\phi
)+{1\over 2}m_3^2\phi^\dagger\phi +{1\over 24}\lambda_3(\phi^\dagger\phi
)^2\nonumber\\
&
-{1\over 4\pi}(m_D^2+{1\over 4}g_3^2\phi^\dagger\phi
)^{3/2}+L_{c.t.}^{II},\nonumber\\
&
\mu=1+{g_3^2\over 24\pi(m_D^2+{1\over 4}g_3^2\phi^2\phi
)^{1/2}},~~~L_{c.t.}^{II}=-{1\over 2}\phi^\dagger\phi ({3\over
2}g_3^2+\lambda_3){\Lambda\over 2\pi^2}.
\label{red2ac}
\end{eqnarray}
Consistent omission of higher dimensional operators requires setting in the
expression of the "magnetic susceptibility" either $\phi =0$ or $\phi
=\phi_{min}$, while the non-analytic piece of the potential should be expanded
into power series of $\phi^\dagger\phi$ up to the ${\cal O}((\phi^\dagger\phi
)^2)$ term. The final 3-dimensional Higgs model is characterised by the
modified couplings
\begin{equation}
\bar m_3^2=m_3^2-{3g^3\over 16\pi}\sqrt{{5\over 6}}T^2,~~
\bar\lambda_3=\lambda_3-{27g^3\over 160\pi}\sqrt{{5\over 6}}T,~~
\bar g_3^2=g_3^2-{g_3^4\over 24\pi m_D}.
\end{equation}
(In the present form of the effective couplings, expansion around
$\phi^\dagger\phi =0$ is assumed.)

An attempt has been made to go beyond the 1-loop accuracy in the integration of
$A_0$ \cite{patkos}. The evolution of the joint potential of $A_0$ and $\phi$
has been followed as the upper limit on the momenta of the Fourier expansion of
$A_0$ has been lowered, with help of an "exact" renormalisation group equation.
Technically, the dependence of the potential on $A_0$ has been made formally
quadratic with the introduction of an auxilliary field $\chi$. Assuming
constant (low frequency) background for $A_0,~\phi ,~\chi$ one can integrate
over the infinitesimal momentum layer of $A_0$. In order to determine the
functional dependence of $\chi$ on $A_0$ and $\phi$, the resulting expression
of the potential energy term of the action is extremised in $\chi$. The
following differential change of the potential energy is found:
\begin{equation}
k{\partial U_k(\phi ,~A_0)\over \partial k}=-{3\over 4\pi^2}k^3\log
(k^2+m_D^2(T,k)+\chi [A_0,\phi ]_k+g_3^2(T,k)\phi^\dagger\phi ),
\label{rgeq}
\end{equation}
where $m_D^2(T,k),~g_3^2(T,k)$ are the coefficients of $A_0^2/2$ and of
$A_0^2\phi^\dagger\phi /2$, at the actual upper momentum scale $k$,
respectively. Their expressions appearing in (\ref{3dcoup}) are the initial
values for the integration of (\ref{rgeq}), imposed at $k=\Lambda$.

The couplings of the effective Higgs theory are obtained when one ends the
integration of (\ref{rgeq}) at $k=0$. Assuming that $m_D^2(T,k)$ deviates from
$m_D^2(T,\Lambda )$ only in higher orders of $g_3$, and $\phi$ fluctuates on
scales smaller than ${\cal O}(T)$, one can justify the expansion of both sides
of (\ref{rgeq}) into a polynomial expression of the fields. Truncating the
infinite coupled set of differential equations at dimension 4 operators, one
finds the following approximate solution for the correction of (\ref{3dcoup}):
\begin{eqnarray}
&
\bar m_3^2-m_3^2=-{3g^3\over 16\pi}\sqrt{{5\over 6}}T^2(1+\sqrt{{6\over
5}}{5g\over \pi^2})(1+\sqrt{{6\over 5}}{17g^3\over 128\pi^3})^{-1},\nonumber\\
&
\bar\lambda_3-\lambda_3=-{27g^3\over 160\pi}\sqrt{{5\over 6}}T(1+\sqrt{{6\over
5}}{17g^3\over 128\pi^3})^{-1},\nonumber\\
&
L_{c.t.}^{II}=-{1\over 2}\phi^\dagger\phi({3\over 2}g^2+\lambda-\sqrt{{5\over
6}}{15\over 32\pi}g^3){\Lambda T\over 2\pi^2}.
\end{eqnarray}
The variation seen in $\bar m_3^2-m_3^2$ and in $L_{c.t.}^{II}$ arising from
the 3-dimensional running of the couplings, driven by the integration over
$A_0$, is about 15\% (for the realistic range of the gauge coupling values:
0.5-0.7) relative to the result of the 1-loop integration over $A_0$.

\section{Lattice investigation of the effective 3D Higgs model}

The lattice Higgs action with general scalar self-interaction potential $V_4$
is written in terms of field variables, appropriately scaled by the lattice
spacing, as follows:
\begin{eqnarray}
&S[U_{xi},\psi_x]=\sum_x[{1\over 2\kappa}\psi_x^\dagger\psi_x-{1\over 2}\sum_i
(\psi_x^\dagger U_{xi}\psi_{x+i}+\psi_{x+i}^\dagger U_{xi}^\dagger\psi_x)]
\nonumber\\
&
+\sum_xV_4(\psi_x^\dagger\psi_x)+S_{gauge}
\label{latac}
\end{eqnarray}
The hopping parameter $\kappa$ is related to the quadratic part of the bare
action through the equality:
\begin{equation}
{1\over 2\kappa}={1\over 2}m_R^2a^2+{1\over 2}({3\over 16}g^2+{1\over
12}\lambda-{3g^2m_D\over 16\pi T})\Theta^2-\Theta\Sigma (N^3){f_{1m}\over 2}+3,
\label{hopeq}
\end{equation}
with $\Theta\equiv aT,~f_{1m}=3g^2/2+\lambda ,~\Sigma (N^3)=\sum_n
(4N^3\sin^2(\pi n_i/N))^{-1}$, $a$ is the lattice spacing, N is the lattice
size.

For any finite value of $\Theta$ one can study the phase transition occuring in
(\ref{latac}) at $\kappa =\kappa_c(\Theta )$. Under the assumption that the
effective action is correctly representing the finite temperature system up to
arbitrary high spatial momenta one takes the limit $a\rightarrow 0$ and
determines the nontrivial limiting quantity:
\begin{equation}
Z_c=\lim_{\Theta\rightarrow 0} ({1\over 2\kappa_c(\Theta)}-3+\Theta\Sigma
(N^3){f_{1m}\over 2}){1\over \Theta^2}.
\label{conteq}
\end{equation}
Using $Z_c$ in (\ref{hopeq}) one finds the physical value of the transition
temperature in proportion to the renormalised mass parameter $m_R^2$ (or the
Higgs mass):
\begin{equation}
{m_R^2\over 2T_c^2}=Z_c-{1\over 2}({3\over 16}g^2+{1\over 12}\lambda
-{3g^2m_D\over 16\pi T}).
\end {equation}

Relation (\ref{conteq}) has been analysed very carefully for a model of the
electroweak phase transition with all variables integrated out on 1-loop level,
but the Higgs-dublet \cite{karsch}. Careful quantitative analysis has shown
(with the Higgs mass chosen around 35 GeV) that $\kappa_c^{-1}$ actually
follows quadratic dependence on $\Theta$ in the interval $\Theta\in 0.1-1.0$.
Detailed finite size scaling analysis of the lattice data for $\Theta\in 1.-3.$
have proven that the transition is discontinous. Monte Carlo estimates of the
latent heat, order parameter jump, etc. are easy to translate into physical
units once the transition temperature is known.
However, the results for the order parameter discontinuity, the latent heat,
and especially for the interface tension were systematically underestimating
the results obtained in simulations of more complete representations of the
electroweak theory \cite{kajantie,csikor}.
One reason for this is certainly the oversimplified perturbative treatment of
the magnetic vector fluctuations (though magnetic screening has been accounted
for by introducing the corresponding screening length into the scalar model in
analogy to the Debye mass, cf. (\ref{red2ac}), by hand). Another question is
whether it is correct to expect that the {\it exact} continuum limit of the
cut-off effective theory should provide the most faithful representation of the
original finite temperature theory.

\section{More accurate reduction: non-renormalisable and non-local operators in
the effective action?}

Higher dimensional operators appear already in 1-loop reduction upon expanding
the fluctuation determinant in higher powers of the background fields. In
\cite{jako} the strength of all non-derivative dimension 6 operators has been
extracted and small numerical coefficients gave argument for their consistent
omission. Even though, if these operators are included into the solution of the
effective model they contribute a linearly divergent piece to the couplings of
dimension 4 operators (we give the expression in lattice regularisation):
\begin{eqnarray}
&
\Sigma{1\over \Theta}\{-[(A_0^a)^2]^2{\zeta (3)g^6\over 16\pi^4}-
(A_0^a)^2\phi^\dagger\phi{\zeta (3)g^2\over 2048\pi^4}[{967\over 2}g^4+{47\over
3}\lambda g^2+{5\over 3}\lambda^2]-\nonumber\\
&
-(\phi^\dagger\phi )^2{\zeta (3)\over 1024\pi^4}[{255\over 16}g^6+{65\over
2}\lambda g^4+{23\over 3}\lambda^2g^2-{40\over 9}\lambda^3]\}.
\end{eqnarray}

 Since these operators are of ${\cal O}(g^6,\lambda g^4, ...)$, the
corresponding "counterterms" will be induced at 3-loop level of the reduction.
 If one does not include into the expected $\Theta$-dependence of the
corresponding bare couplings these contributions, a theoretical error is
introduced into the reduced description. In the $\Theta$ region where the terms
with inverse $\Theta$-dependence are negligible the error is negligible.
Clearly this requirements (one for the coupling of each dimension 4 operator)
sets a lower limit to the variation of $\Theta$. When one substitutes the usual
numerical range for $g$ and $\lambda$ it turns out that these lower limits for
the grain size are ${\cal O}(10^{-2})(1/T)$. This makes the preceding
discussion only of conceptual interest, since these limits represent a warning
that one should not take for any of the approximate reduced models the strict
continuum limit.

Effects of two-loop level reduction in the 3 dimensional representation of
finite T field theories has been discussed on the example of the N-component
scalar theory \cite{jako3}:
\begin{equation}
S=\int_{0}^\beta d\tau\int d^3x[{1\over 2}(\partial_\mu\phi_\alpha )^2+{1\over
2}m^2\phi_\alpha^2+{1\over 24}\lambda(\phi^2_\alpha )^2].
\end{equation}
The coefficients $m^2_3(T),\lambda_3(T)$ have been explicitly determined:
\begin{equation}
m_3^2=m^2+({1\over 24}\lambda -0.001355\lambda^2){N+2\over
3}T^2,~~\lambda_3=\lambda T.
\label{scalredpar}
\end{equation}
Counterterm has been induced only to the mass of the 3 dimensional theory:
\begin{equation}
\delta m_3^2=-{N+2\over 3}[{\lambda\over 4\pi^2}(1-0.048277\lambda )\Lambda
T+{\lambda^3\over 32\pi^4}\Lambda T\log{\Lambda\over T}+{\lambda^2\over
48\pi^2}T^2\log{\Lambda\over T}]
\label{divcontred}
\end{equation}
(again cut-off regularisation has been used).

A puzzling observation is made when one calculates at 2-loop level the
effective potential of the effective theory with parameters taken from
(\ref{scalredpar}). The divergent part of the potential turns out to be
\begin{equation}
U_{div}(\phi_0)={1\over 2}\phi_0^2{N+2\over 3}[{\lambda\over 4\pi^2}\Lambda
T-{\lambda^2\over 192\pi^2}T^2\log{\Lambda^2\over \mu_3^2}].
\label{div3dpot}
\end{equation}
This result agrees with the scalar part of the 2-loop calculations done for the
3 dimensional SU(2) Higgs model by several authors
\cite{farak2,laine,kripfganz}. There is no doubt that one encounters a mismatch
between all 3 types of divergences occuring in the 2-loop mass counterterm and
the
divergent piece of the effective potential they should cancel. One should
interpret this as a signal that the representation of the finite T theory with
an effective action containing local operators up to dimension 4 cannot be
correct at 2-loop level. By the previous discussion also the possible impact of
higher dimensional local operators can be ruled out, since they are of higher
order in $\lambda$.

The resolution proposed in \cite{jako3} is to include also a momentum dependent
4-point vertex into the effective theory. This operator is constructed to
reproduce those contributions to the effective potential of the full finite
temperature theory, which diagrammatically contain both static and non-static
internal lines. Such situation occurs first at 2-loop level, what explains why
at 1-loop no mismatch has been observed. For the N-component scalar model the
non-local term was found to be of the following form:
\begin{eqnarray}
&
L_{non.loc.}^{3D}={\lambda^2\over 128\pi^2}({N+4\over 9}O_1+{4\over 9}O_2),
\nonumber\\
&
O_1=\phi^2_\alpha\Omega (i\partial )\phi^2_\beta
,~~O_2=\phi_\alpha\phi_\beta\Omega (i\partial )\phi_\alpha\phi_\beta,
\nonumber\\
&
\Omega (k)={\pi^2Tk\over T^2+k^2}+{1\over 2}\log (1+{k^2\over
4C^2T^2})+F({k\over \Lambda}),
\end{eqnarray}
with $C=2\pi\exp (1-\gamma_E)$ and $F(x)=(2/x-1)\log (1-x/2)+1.$

Since $L_{non.loc.}^{3D}$ is ${\cal O}(\lambda^2)$ for the solution of the
effective theory to ${\cal O}(\lambda^2)$ it is sufficient to calculate its
contribution at 1-loop. Then one can check explicitly that its divergent
contribution exactly fills the gap between (\ref{div3dpot}) and
(\ref{divcontred}). This means that at least to two-loop accuracy there is
consistency between the full theory and its reduced image up to arbitrarily
high spatial momenta.

If one is not interested in momenta above the scale ${\cal O}(T)$, the effect
of the non-local terms would be compressible into the couplings of a local
effective $\phi^4$-type theory of the effective theory. Also the field
variables of the two theories might differ, therefore one should introduce a
field rescaling factor Z:
\begin{equation}
L_{eff,eff}={1\over 2}(\partial_i\hat\phi_\alpha )^2+{1\over 2}\hat
m_3^2\hat\phi_\alpha^2+{\hat\lambda\over 24} (\hat\phi_\alpha^2)^2,~~
Z\phi_\alpha^2=\hat\phi_\alpha^2.
\label{loceff}
\end{equation}
For the determination of the couplings one can follow the matching strategy of
\cite{braaten} and require the equality of the 2- and 4-point functions
calculated from (\ref{loceff}) and from the non-local theory at small spatial
momenta.
Systematically throwing away terms of ${\cal O}(m_3^2/T^2)$ one finds, for
instance:
\begin{eqnarray}
&
Z=1-{(N+2)\lambda^2\over 96\pi^2}({1\over 6}\log{T^2\over m_3^2}+{1\over
2}+{1\over 12\pi C}+{\cal O}({m_3^2\over T^2},~\lambda )),\nonumber\\
&
\hat m_3^2=m_3^2(T)+{N+2\over 3}\lambda^2T^2({C\over 32\pi^3}+{\cal
O}({m_3^2\over T^2}\log {T^2\over m_3^2},\lambda )).
\end{eqnarray}

Actually, the matching strategy of the last step has been applied directly to
finding the coupling parameters of the local 3 dimensional Higgs model
representation of the finite temperature SU(2) Higgs system \cite{farak2}. The
2-loop effective potential of the effective theory has been computed. In this
superrenormalisable theory only the mass needs renormalisation and the constant
characterising its scale dependence has been determined by comparing the result
to the 2-loop calculation of the effective potential in the full model. For the
interpretation of the lattice investigations of this particular effective model
the analogue of the relation (\ref{hopeq}) has been carefully determined to
2-loop accuracy \cite{farak3,laine2}. In view of the above discussion it cannot
be taken granted that the scaling behavior of the local effective theory, valid
in that model for very large values of the cut-off, can be observed for
$aT\sim{\cal O}(1)$, where this theory is expected to describe the original
finite temper
ature theory well. The observation of a scaling window around $\Theta\sim 1$
seems to be a rather non-trivial task.
Even if it can be observed for a specific discretisation, one ought to check
the compatibility of the results obtained with different discretisations (in
order to check the absence of finite lattice spacing effects).

\section{Acknowledgements}

I wish to thank my collaborators A. Jakov\'ac, K. Kajantie, F. Karsch, T.
Neuhaus, P. Petreczky and J. Polonyi and also a grant by the Svenska Institutet
(No. 272/399/34) allowing me to enjoy the hospitality of the University of
Uppsala, where this contribution has been completed.

\end{document}